\documentclass[galaxies,article,submit,oneauthor]{mdpi} 
\usepackage{pdflscape}
\firstpage{1} 
\pubvolume{1}
\issuenum{1}
\articlenumber{0}
\pubyear{2021}
\copyrightyear{2020}
\datereceived{} 
\dateaccepted{} 
\datepublished{} 
\hreflink{https://doi.org/} 
\pdfoutput=1

\Title{Somewhere in between: tracing the radio emission from galaxy groups (or why does the future of observing galaxy groups with radio telescopes look promising)}

\TitleCitation{Title}

\Author{B\l a\.zej Nikiel-Wroczy\'nski $^{1}$\orcidA{}}

\AuthorNames{B\l a\.zej Nikiel-Wroczy\'nski}

\AuthorCitation{Nikiel-Wroczy\'nski, B.}

\address{%
$^{1}$ \quad Astronomical Observatory, Jagiellonian University, ul. Orla 171, PL 30-244 Krak\'ow, Poland; blazej.nikiel\_wroczynski@uj.edu.pl}

\corres{Correspondence: blazej.nikiel\_wroczynski@uj.edu.pl; Tel.: +48-12-6238647}

\abstract{Galaxy groups constitute the most common class of galaxy systems in the known Universe, unique in terms of environmental properties. However, despite recent advances in optical and infrared observations as well as in theoretical research, little is known about magnetic fields, and the associated continuum radio emission. Studies on this issue have only been conducted in recent years, and many questions have yet to be resolved. This article aims to put the study of group magnetism in a broader context, to present recent advances in the field (mainly achieved with low-frequency radio interferometers), and to list the issues that need to be addressed in future observations. To make it easier for the Readers to get acquainted with the concepts presented in the manuscript, radio observations of two sample groups of galaxies are also presented. }


\begin{document}
\section{Introduction: galaxy groups as a very special class of objects}

Galaxies in the Universe live either in solitude, or in interacting systems. The former ones are said to be a mere exception; as \citet{klein18} (and many other beforehand) argue, most of galaxies are bound to other ones by gravitational force. There are three main classes of galaxy systems: pairs, groups, and clusters. Pairs are the easiest to describe: it is just a duo of interacting galaxies, of moreover similar sizes and masses. M51 and its companion, NGC 5195, are probably the best known example of a pair -- however, due to a large difference between the larger and the smaller companion, they are not the best one though. Taffy galaxies -- systems of interacting spirals, one detected by \citet{condon93}, another one by \citet{condon02} -- are a much better illustration of what a pair of galaxies looks like. Both of these systems contain not only member galaxies, but also a wide, magnetised intergalactic bridge. Such structures are among the main drivers of studies on interacting systems.

Compared to a pair, a cluster would be a totally different object. They are the largest, "homogeneous" galaxy systems, as superclusters are objects composed of several different clusters and/or groups. Unlike lowly pairs, clusters include hundreds, if not thousands of members, all immersed in a Mpc-wide halo, visible well in the radio and X-Ray regimes. Being quite rewarding targets, massive clusters of galaxies have attracted a lot of attention, and are frequent protagonists of astronomical studies, including those made in the radio continuum (see eg. the newest review by \citealt{vanweeren19}). 

Groups that constitute the middle tier of galactic systems  the subject of scientific research about 40 years ago. Even so, many aspects of group evolution, composition, or environmental characteristics remain scarcely studied. Among them is the presence and nature of the magnetic fields within -- one of the issues that are crucial to understand the physics of the groups' intergalactic medium, and its members.

Why are magnetic fields expected to be that important? A detailed answer can be found in \citet{beck13}, who compiled a long lists of processes where they play a profound role. Magnetic fields regulate the process of stellar formation, as they can both quench (by stabilising the gas clouds), and give an onset to the latter (by removal of the angular momentum through the ambipolar diffusion). Magnetic fields contribute to the total pressure within galactic disks, effectively balancing them, and can become a source of heating and turbulence in the outer disks of spiral galaxies. Even more: magnetic reconnection can heat both the interstellar medium, and the gas within the galaxy's halo. Magnetic fields can serve as a constraining mechanism for the creation of shocks in density waves. And finally, as the energy contained within the magnetic field is comparable to the thermal one, the dynamics of the gaseous flows cannot be understand without magnetism being taken into account. And all of these phenomena are likely to happen in groups, where studies on the magnetic fields have just begun. But before this issue can be addressed, it is advisable to recall the basic information on this class of galaxy systems. 

The first galaxy group to be described was the Stephan’s Quintet, named after its discoverer, French astronomer Eduard Jean-Marie Stephan, who reported a small grouping of nebulae in the constellation of Pegasus \citep{stephan77}. However, no special attention has been paid to them for more than 100 years. An important milestone in understanding the importance of systems of interacting galaxies was the first n-body simulation in astrophysics, done by Erik Holmberg \citep{holmberg41}. Holmberg wanted to investigate how does a close passage of two galaxies affect them. Using a set of bulbs with illuminance detectors (because illuminance scales with distance in the same way as the force of gravity), assuming initial parameters such as speed or direction, he did the same as computers do now: step by step he moved individual "particles" according to these conditions and calculated the gravitational pull.  This \emph{brilliant} project led to the groundbreaking discovery that galaxies in systems interact in a seemingly destructive way, with the initial spiral shape being elongated into tidal structures, and finally, completely wiped out in favour of a moreover structureless body. This was the moment when it became clear that interactions must be studied in detail.

However, Holmberg’s discovery did not spawn myriads of scientific publications on interacting systems. In case of groups, there were several case studies: Stephan's Quintet in particular received a lot of attention, but other groups were not as popular.The earliest study that focuses primarily on compact galaxy groups is that of Shakhbazyan -- actually, a series of ten papers written by different authors, from 1973 to 1979 \citep{shakhbazyan73}. Shakhbazyan noted that in addition to "normal" clusters, compact clusters and groups, consisting of compact galaxies are also observed -- and these entities are not rare, if one takes faint galaxies into the account.  Moreover, these compact structures are unlikely to be ejected subsystems of larger clusters; they are more likely to form independently of clusters, especially since many isolated, compact groups of galaxies have been discovered.  While the original paper referenced 30 systems, the complete list is much larger and consists of 377 different systems, from triplets to those comprised of dozens of galaxies. Other examples include the catalogue of groups and clusters by \citet{klemola69}, Zwicky’s discussion of the tidal tail of the Leo Triplet (\citealt{zwicky56}, followed by Kormendy and Bahcall’s paper of 1974, \citealt{kormendy74}) or Arp's catalogue of interacting galaxies from 1966 \citep{arp72}.

The person to change the way we perceive galaxy groups was Paul Hickson, who prepared a set of three criteria for a system to be regarded as a compact group –- quantity, compactness, and isolation –- leading him to the creation of a 100-object sample, known today as the Hickson’s Compact Groups \citep{hickson82} -- which included both systems previously described by \citet{shakhbazyan73} and other Authors, as well as newly discovered ones. The Palomar Survey’s optical data were enough to notice that galaxies found in these systems do not resemble those found in the field. Effects of interactions were evident: members showed morphological and kinematic peculiarities, infrared and radio emission from their nuclei has already been found, they seemed to host large amounts of diffuse gas and dark matter. Their evolution appeared to be controlled by interactions, with the final step being the ultimate demise of the system in question. It was also reported that groups were “surprisingly numerous, and may play a significant role in galaxy evolution” -- later backed up by \citet{mulchaey00} who concluded that not only are groups numerous, but they might even comprise the most typical environment in the Universe.

Hickson’s findings marked the beginning of an era of studying groups. His sample was evaluated by many, searching for signs of different processes. It’s completeness was also discussed; as the original study does not contain any redshift information, Hickson revised it in 1992 \citep{hickson92}, ruling out five systems (9,11,36,41,77) as not being groups at all; in addition, some of the groups do not meet the criteria any more (the best example is the Stephan’s Quintet, catalogued as HCG 92, in case of which galaxy NGC 7320 is just a foreground interloper; without it, it should not be regarded as a HCG). However, arguments in favour of the real character of such compact groups have also been presented, by eg. \citet{tovmassian99}. What has been learned about groups so far? Galaxies in groups are very special ones. An analysis of their infrared parameters, their star formation rate, their position on the colour-colour diagram –- in comparison with that from clusters, or field ones -– reveals a very specific, bi-modal distribution \citep{walker10, walker12, walker13}: group members are either dim, and weakly form stars, or are they are surprisingly bright, undergoing a starburst phase. There is a lack of what could be called “normal” galaxies: forming stars at a modest rate. No such occurrence was reported in the control samples: one taken in the centre of the Coma cluster, another one -- consisting of the field galaxies. Only a mild correspondence could be found between galaxies in groups and in the Coma “inflow region”, where galaxies are sucked into the cluster. This was interpreted by those Authors as an obvious sign of an accelerated evolution of galaxies in groups, and –- given the fact that this phenomenon was clearly detected only in HCGs -– attributed to the environment. These results were backed by the previous knowledge of physical conditions in groups: while a few HCGs show signs of “typical” cluster phenomena, like diffuse starlight, severe H$_{\rm I}$ depletion, dominance of “red and dead” galaxies , many are still “blueish”, undergoing a starburst event \citep{huchtmeier97}. 

Another important study exploring the uniqueness of the group environment is that of \citet{paul17}. These authors analyse a mock catalogue of systems of galaxies that differ in mass, radius, and temperature. Simulations reveal that while in the “upper” subset –- the division line is at a mass of 8 × 10$^{13}$\,M$_{\odot}$, a temperature of 1 keV, and a radius of 1 Mpc –- self-scaling laws for clusters do apply, it does not work this way in the “lower” one –- systems that resemble galaxy groups. Therefore, it can be clearly inferred that galaxy groups are indeed special environments: there is probably no evolutionary similarity between galaxies inside and outside of them. And since the groups have already been introduced as the most common environment in the universe, it can in fact be assumed that the "special" evolutionary path is actually "normal" and vice versa.

There exists also a very special subtype of groups: fossil groups, or simply fossils. They were proposed as a late evolutionary stage of a small cluster. Subsequent acts of galactic cannibalism lead to the formation of a large, elliptical galaxy \citep{hausman78}. When this happens in a group -- as shown by \citet{barnes89} -- the result is the same: only a single, bright elliptical remains (or this + several other members, if the system is "caught" at an earlier stage). The fossil is therefore a "dead" group, opposed to the normal ones, where merging processes, and interaction-driven phenomena related to them are still ongoing. However, a fossil can be reborn: there is a possibility that the elliptical galaxy in its centre will serve as a gravitational attractor, and will create a new system by stripping nearby groups and clusters from their members. The first detection of a fossil group is attributed to \citet{ponman94}, who found a promising candidate at a redshift of 0.171: a galaxy comparable to brighter HCGs with regards to its X-Ray emission, yet apparently isolated.

What are the characteristic traits of fossil groups? They are expected to be comprised of a single early type galaxy, or a few ones, immersed in a large X-Ray halo -- remnant of the previous interactions. Such haloes have become the hallmarks of fossils: while plain and uninteresting with regards to their optical images, systems like NGC\,6482 \citep{khosroshahi04}, or HCG\,15 \citep{mulchaey03} are real "eye-catchers" when viewed by an X-Ray observatory. The growing interest in studying fossils relates to their special evolutionary status: they are at a well localised -– on the timeline of the process –- moment of a group’s evolution. As groups are expected to evolve via interactions \citep{hickson82}, one can easily identify a fossil as a probable evolutionary stage in lives of most of the groups. Still, not much is known about the physical properties of them. For example, while simulations usually suggest that a fossil is a relaxed system, \citet{zarattini16} argue that “real” fossils can still have subsystems –- is this because there are different evolutionary tracks for fossils, or because those “real” ones are systems prior to final merging? If indeed being a fate to be faced by most of the groups, why are fossils so rarely observed -– is it a matter of a short temporal span of the fossil stage (so either objects at several steps prior to this one, or systems “reborn” are seen), or rather an indication that some systems might finish their evolution at an even earlier stage? Or maybe some groups and clusters simply start aggravating more mass even before the final merging event, so at the time the central galaxy forms, it already has quite a large family? 

But whether it's a mysterious fossil or a better-studied living group of galaxies, little is known about the magnetic fields there. It comes with a twist; the magnetic field has already been mentioned as one of the most important factors in galactic life. It is usually revealed by the synchrotron effect, resulting in radio continuum emission. To effectively fuel this process with cosmic rays (CRs), vigorous star formation is necessary: in galaxies, CRs are provided mostly by the supernovae, which mark the end of an evolution of the rarest, most massive stars -- so groups, which are known for their starbursting member galaxies \citet{hickson82}, should be then an easy target. However, revealing this phenomenon turned out to be a nightmare: a group is a tricky object to be observed by a radio telescope.

How is it even possible? First of all, groups are rather small objects, and only the nearest ones are angularly large; this is why single-dish radio telescopes are almost never used to study them. Interferometers can be more of an use here; however, what one is searching for are extended, rather weak structures. In order to achieve a high resolution -- some 10--20 arcseconds -- sparse antenna distribution is needed, translating into a decreased sensitivity to the extended emission. This was especially true at lower frequencies, were either the beam was too large, or the sensitivity was too low to conduct a magnetism study on group scales at all. One might then assume that research at higher frequencies could be the solution. But here comes another important problem: stars that end up as supernovae are easy to find in galactic disks (or even in some detached, compact starburst regions), but generally not in cooler gas tails and bridges through which the magnetic field can be stripped from their parent galaxies.The higher the emission frequency, the faster the emitting particles disappear. For this reason, even bright galaxies are known to have their halos fade away quickly with frequency, to the point where emission is dominated by more compact thermal regions. And these unfavourable limitations are the reason why only a few case studies of radio emission and / or magnetic fields in galaxy groups have been made over many years. 

Fortunately, a lot has changed in the past few years, with the introduction of the new low frequency radio intereformeters, and modernisation of the existing ones. Instruments like the International LOFAR Telescope (ILT, \citealt{vanhaarlem13}), the Murchinson Widefield Aray (MWA, \citealt{MWA}), or the Upgraded Giant Metrewave Telescope (uGMRT, \citealt{uGMRT}) have greatly contributed to making the research of intergalactic magnetic fields possible. And the future looks even brighter, with the planned update of the ILT (LOFAR 2.0), or the soon-to-be-built Square Kilometre Array (SKA), the biggest radio telescope in the world. 

This paper is designed as a longer version of the presentation entitled "Somewhere in between: tracing the radio emission from galaxy groups", delivered during the \textit{A new window on the radio emission from galaxies, clusters and cosmic web} conference, on 9th March of 2021. It recapitulates the current state of knowledge about galaxy groups and their magnetic fields and, while waiting for new discoveries, identifies questions that still need to be answered, and provides evidence that such research will indeed be possible with current infrastructure developed.

\section{Materials and Methods}

In order to provide example imaging material demonstrating the quality of the data one can obtain using the low-frequency infrastructure, two galaxy groups have been chosen: HCG\,15 (a radio-emitting fossil), and HCG\,60 (an "alive" galaxy group containing a radio galaxy). In case of the latter, data from LOFAR, GMRT, and VLA have been included. The LOFAR ones come from the DR1 cutout service system (available at https://lofar-surveys.org/releases.html). They were acquired already calibrated, and imaged; thus no further processing was necessary. The GMRT ones -- from \citet{giacintucci11}, at 612\,MHz -- were downloaded from the GMRT Online archive in a raw form, and were then processed using the \textit{Source Peeling and Atmospheric Modeling} software \citep[SPAM,][]{IntemaPhD2009, Intema2009, Intema2014}. SPAM makes it possible to acquire science-ready images with a minimal input. The calibrated data were imaged using \textsc{wsclean} \citep{offringa14}, which is used as a standard imager for the LOFAR data. Primary beam correction was applied using the coefficients provided by N. Kantharia. Archive VLA data of \citet{menon85}, at 1550\,MHz, were obtained in raw format, and processed in accordance with the standard procedure outlined in the NRAO Astronomical Image Processing System (\textsc{aips}) Cookbook. 

In case of HCG\,15, GMRT and VLA data were used. The latter ones -- at the frequency of 4.86\, GHz -- were taken from \citet{bnw17}. Processing of the GMRT data (this time, at 618 MHz) was the same as in case of HCG\,60, while the VLA ones were acquired in their final form (calibrated and imaged).

Images from both of these instruments were loaded into \textsc{aips}, convolved to a common resolution of 6 arcseconds in case of HCG\,60, and 18 arcseconds in case of HCG\,15. The geometry of those images has been conformed to that of an appropriate, POSS-II red plate used as an optical background.

\section{The character of the radio emission from galaxy groups: current knowledge, and future perspectives}
\label{sec_general}

Radio studies of galaxies in groups, and groups themselves date back to the 70's of the XXth century. It comes with a little to no surprise that the first object to be studied was the Stephan's Quintet, by \citet{allen72}, who used the freshly built Westerbork Synthesis Radio Telescope (WSRT). Apart from an unresolved source of radio emission, close to the nucleus of NGC\,7319, an unexpected discovery was made: the resolution of 25 arcseconds was enough to detect an extended source of radio emission in a form of a semicircular arc, partially encompassing the aforementioned galaxy. Thus, a first detection of an extended radio emission from the IGM of a galaxy group was made.

While the Quintet was revisited shortly after by other researchers (e.g. \citealt{kaftankassim74, kaftankassim75, gillespie77, vonkapherr77}), it was not before the Hickson's paper on compact galaxy groups \citep{hickson82} that an attempt to detect radio emission from a larger sample of these objects was undertaken. In 1985, \citet{menon85} did a survey of 88 compact groups using the VLA, at the frequency of 1550\,MHz, utilising a rather sparse, B-configuration of the array. The Quintet -- which is included in the said catalogue as HCG\,92 -- was purposefully omitted, as its radio emission was already detected. The results of the study were not encouraging, however. Only one group has been found to host an intergalactic, radio emitting structure –- HCG 60. It  even looked like most of the galaxies in groups are radio-quiet ones: in total, only 33 galaxies were detected, in just 11 groups. The success ratio was therefore around 14\% for detecting a group containing a radio emitting galaxy, and around 2\% for the detection of an extended emission. No attempt to take measurements of the magnetic fields were carried out.

Discouraging results of this project were one of the reasons why for more than thirty years, there was a very limited progress in studying the continuum radio emission from galaxy groups. As already stated in the introduction, it is not an easy target for observations -- thus, the overall probability of success of such studies is rather low. A few case studies have been successful -- eg. that of Ho124 \citep{kantharia05}, where an intergalactic structure connecting all members of this interacting triplet was detected. The Quintet was visited again, twice. The first study was done by \citet{xu03}, who used high resolution VLA data to study the "arc" of \citet{allen72} -- actually, a large intergalactic shock -- as well as made the first discovery of a radio emission from a tidal dwarf galaxy (TDG) candidates SQ-A and SQ-B. The second study was that of \citet{bnw13B}, who focused on the extended sources of radio emission. System’s halo was seen with an unprecedented fidelity, both at 1.4 and 4.86 GHz; it turned out that this structure is not an effect of smearing out the contribution from the shock region seen in details by \citet{xu03}; instead, member galaxies NGC 7318A, 7318B and 7319 are all immersed in a genuine, large-scale, radio emitting structure. Hints for the presence of a regular magnetic field were also present (this issue is covered in details in Sec.~\ref{sec_regfield}). Analysis of the strength of the field, and the density of energy contained within – and its comparison with the literature information –- yielded an interesting conclusion: not only was the magnetic field found in the IGM of the Quintet similar in strength to that found \emph{inside} galaxies, but its energy density was comparable to that of the thermal electrons. This was a clear indication that magnetic field can be an important agent in the evolution of a compact group –- something that was expected, as it was already noticed in case of interacting pairs \citep{drzazga11}, but has never been proven before. Another studied object was the Leo Triplet \citep{bnw13A}; a loose group, known for the presence of a more than a hundred-kpc-long neutral gas tail \citep{haynes79, stierwalt09} that accompanies a weak optical tail \citep{kormendy74}. Density in such systems is much lower, so interactions can be less frequent, and limited to eg. a pair of galaxies, with the third one being left quite unscathed for a long time. Tidal structures are therefore more likely to survive for a long time, undisturbed by other group members; these groups are also speculated to be an earlier evolutionary stage of the compact ones \citep{barnes89}. However, in case of the Triplet, only the member galaxies themselves turned out to be radio emitters. The spectropolarimetric study of two southern sky groups by \citet{farnes14} also revealed no signs of an intergalactic, magnetised medium. 

There was still no general picture, however. A survey of 18 radio and X-Ray emitting groups was done by \citet{giacintucci11}, but the emission traced there was mostly due to the AGN activity, not the action of intra-group magnetic fields. An attempt to recapitulate the already gathered cases and evidence was done by \citet{bnw17}. Four objects were reviewed – HCG 15 (from \citealt{giacintucci11}), 44, 60 (the only detection of \citealt{menon85}), and 68. While no detection of an intergalactic, radio emitting structure was done in case of HCG 68, such entities were revealed at 1.4\, GHz in HCG 15 and 60 (in this case, also at 4.86\,GHz), with an additional hint for an intergalactic bridge in HCG 44. Also, all of these systems contained radio emitting galaxies -- a finding much more optimistic than those of \citet{menon85}. Detected intergalactic magnetic fields were comparable to those found inside the galaxies, suggesting that their influence cannot be omitted when studying the dynamics and evolution of galaxy groups. Yet, the number of studied objects was far too low to allow any kind of a statistical study. Fortunately, thanks to the low frequency observations from the LOFAR Two Metre Sky Survey (LoTSS), it finally became possible -- in 2019.

The LoTSS \citep{shimwell17} is designed as the first large survey utilising LOFAR. It is done in the frequency range of 120-168\,MHz, and aims at covering the whole northern sky, in a net of 3170 individual, overlapping fields, each observed for 8 hours. The resolution at the central frequency of 144\,MHz is 6\,arcsec, and the sensitivity level reaches approximately 100\,$\mu$Jy/beam (field-dependent). The observed and processed data are being made available to public in subsequent data releases; the first of them -- outlined by \citet{shimwell19} -- covered the area of the HETDEX Spring Field \citep{hill08}. The public release of the data was accompanied by a "paper splash": a coordinated publication of 26 manuscripts utilising the LoTSS catalogue and imaging products. Among those papers was a study of more than a hundred galaxy groups found within the HETDEX region, the LoTSS of GGroups \citep{bnw19}

The LoTSS of GGroups was intended to serve as a pathfinder project, testing whether sensitive, high fidelity observations at low radio frequencies could overcome the problems usually faced when conducting studies of magnetic field in galaxy groups. A sample consisting of 120 galaxy groups has been prepared by cross-matching systems from the HCG \citep{hickson82} and MLCG (Magnitude Limited Compact Groups, \citealt{sohn16}) with the sky coverage of LoTSS. This included both isolated systems, and embedded ones -- eg. dense cores of inhomogenous clusters that met the isolation criteria of their parent catalogues. The result was more than optimistic: 73 groups contained at least one radio emitting galaxy (a success ratio of 61\%), and 17 of them were hosts of an extended, radio emission – thus, there was apparently a detectable, intergalactic magnetic field in about 15\% of the analysed objects. Four objects -- HCG\,60, MLCG\,24, MLCG\,41, and MLCG\,1374 -- were chosen for a more detailed investigation. Two former were embedded systems, two latter -- isolated ones. Combined data from the NVSS \citep{condon98} and FIRST \citep{FIRST} surveys at 1.4\,GHz were used to supply the information from a second radio frequency (necessary to calculate the strength of the magnetic field). It turned out that in all four galaxy groups, the intergalactic magnetic field has a strength of 6--9\,$\mu$G. This is not only similar to that found inside the non-shocked regions of the Quintet's IGM (6\,$\mu$G -- \citealt{bnw13B}), but also comparable to the magnetic field found inside spiral galaxies (an average of 9\,$\mu$G -- \citealt{niklas95}). From that point it is certain that magnetic fields in galaxy groups can be detected -- and should be investigated, as they are likely to be important agents in shaping their host systems.

The success of this pathfinder project is one of the signs that a new era in the observations of groups is beginning. The sensitivity of the LOFAR data allows to easily detect radio emission from groups and low-mass clusters: since the release of the LoTSS DR1 data, numerous discoveries of an extended radio structure in a lightweight galaxy system have been reported (see eg. \citealt{hoang19, knowles19, botteon19, paul20, botteon21}). Very recently, \citet{paul21} has shown that the uGMRT data gathered at 400\,MHz are complementary to those from the LoTSS and are also a powerful tool to undercover diffuse radio emission in low-mass clusters; detections were made in systems that are characterised by masses around 2--3 times larger, than the theoretical upper limit for galaxy groups modelled by \citet{paul17}. This is a strong hint that the uGMRT can easily handle observing group radio emission, too -- offering not only excellent comparison data, but also a possibility to bring studies on the group magnetism to a whole new level: for example, by using the combined uGMRT and LOFAR data to create high-resolution maps of the strength of the magnetic field within the studied system. Further analysis -- done altogether with the X-Ray data -- can also help to understand where and which energy dominates over another. Analysis of the magnetic field in groups and pairs has also been announced as one of the topics to be covered by the SKA \citet{heald20A}. The excellent capabilities of this instrument should make it possible to study the radio emission from low mass, interacting systems, over a broad range of frequencies.

What are the issues that still need to be addressed, having the results of \citet{bnw19} at hand? First of all, not much is known about the properties of the magnetic field in groups: as for now, it is certain that there are several systems where it is dynamically important, but it is not sure if such objects are frequent, or not; a 6\,$\mu$G-strong magnetic field in one group might contain enough energy to be compared to that of the thermal electrons therein, but that can be not enough in case of another. Is there a typical strength of the intergalactic magnetic field in groups? How does it look compared to the median values found for the intracluster medium and that found inside galaxies? Yet another issue is whether it is even justified to treat all galaxy groups as a single class: the question would be, if there exists any connection between eg. the isolation of systems, and their median magnetic field strength. And maybe there is a clear evolutionary sequence, like in merging pairs \citep{drzazga11}: dense systems host strong magnetic fields, while sparse ones –- do not. A “magnetic evolutionary track” could be revealed that way. Let’s not forget about galaxies, too. From the works of \citet{walker10, walker12, walker13} it is known that the star-forming properties of galaxies in groups are relatively extreme. That should also be reflected in their radio characteristics, and magnetic field -– after all, it’s the star formation that “fuels” the amplification of the former (\emph{it's an oversimplification of events, but yes}). Are group galaxies, in their respective, morphological “bins”, hosting stronger magnetic fields than those outside them? Or maybe there is no visible difference? Cross-matching galaxies from \citep{walker10, walker12, walker13} with the the LoTSS coverage would allow to build the radio counterparts of the samples referenced therein, and make it possible to analyse whether the bi-modal distribution seen in the infrared regime is also present when it comes to the radio luminosity. Last but not least: as far infrared (FIR) data for galaxies in Hickson Compact Groups are already at hand, analysis of the FIR-Radio correlation in galaxy groups is also possible: is the special character of galaxies in groups reflected in a different correlation coefficient? Answers to this whole list of objectives are exactly what is needed to paint the picture of the groups’ magnetism.

\begin{figure*}[htp]
\widefigure
\centering
	\includegraphics[width=0.47\textwidth]{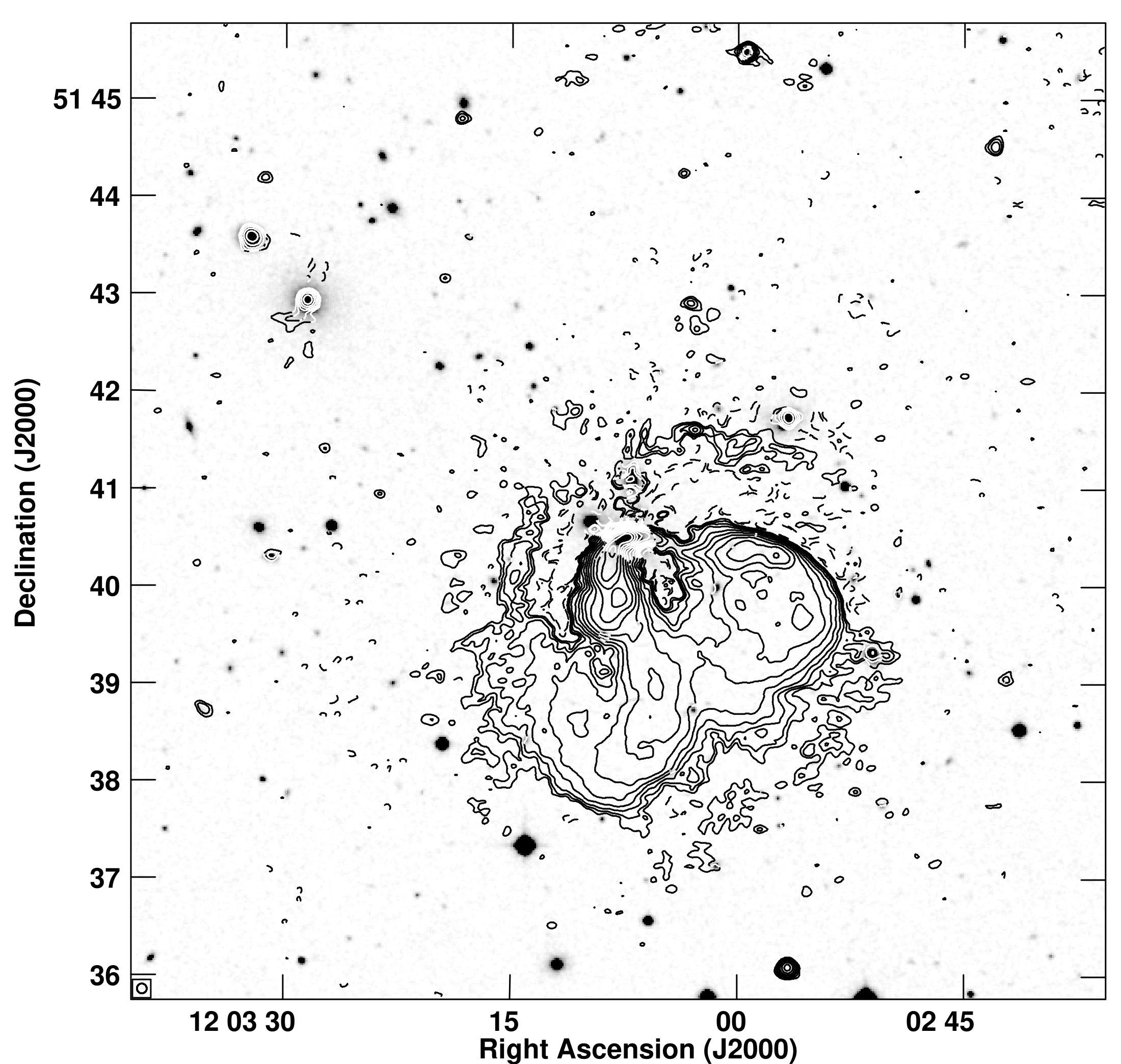}
	\includegraphics[width=0.47\textwidth]{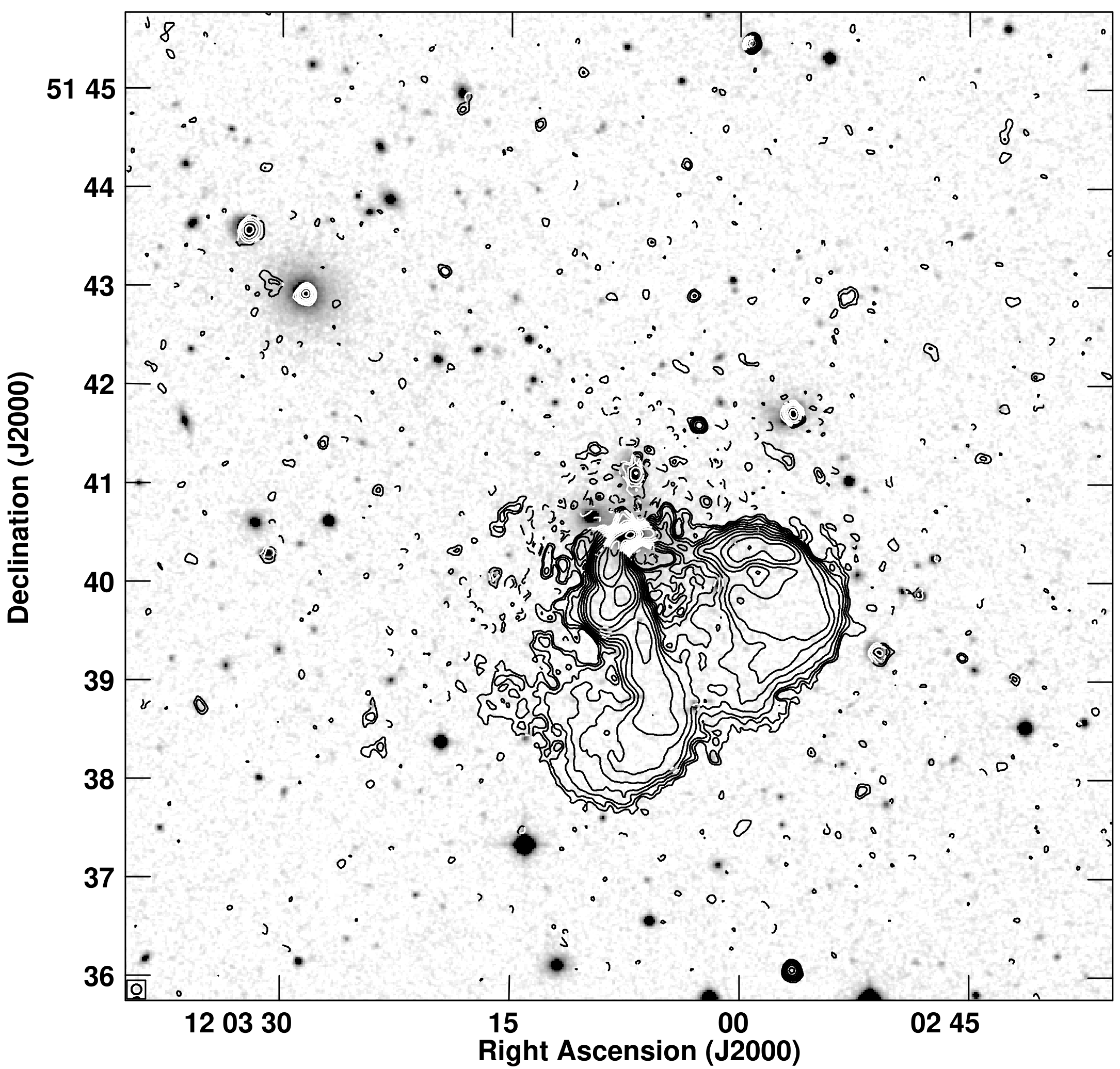}	
	\includegraphics[width=0.47\textwidth]{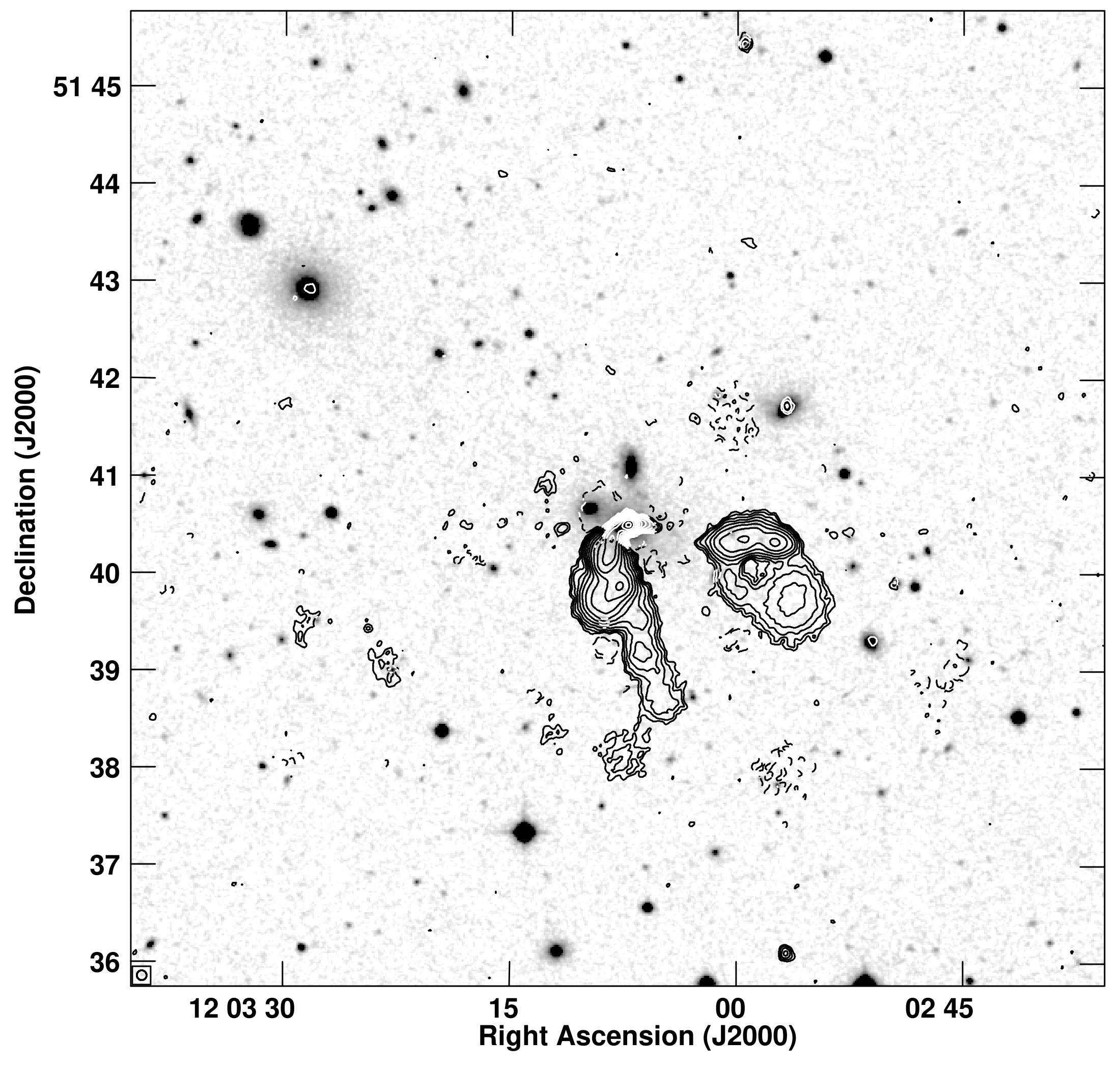}	
\caption{\label{fig_h60} Radio emission of HCG\,60, at 144\,MHz (left panel), at 612\,MHz (right panel), and at 1550\,MHz (bottom panel). Radio contours overlaid on a Johnson R filter map from the DSS. The levels are -5, -3, 3, 5, 8, 13, 21, 34, 55, 89, 134, 213, and 345 $\times$ r.m.s. noise level of 70\,$\mu$Jy/beam (144 MHz), 35\,$\mu$Jy/beam (612 MHz), and 40\,$\mu$Jy/beam (1550 MHz). The common beam is 6 arcseconds.}        
\end{figure*}

To illustrate the capabilities of today's instruments, GMRT observations of HCG\,60 at 612\,MHz were chosen (Fig.~\ref{fig_h60}). While they are not as sensitive as those that can be gathered after the telescope's upgrade, they already show a large pool of radio emission, despite the high resolution of 6 arcsceconds.  They are presented together with the LoTSS DR1 data of \citet{bnw19}, and those from \citet{menon85}, at 1550\,MHz. The other maps also allow to see the details of the radio structure; however, while both low frequency images are complementary to each other, the older, supra-gigahertz data achieve this goal at the cost of sacrificing the sensitivity to extended structures. No traces of the large halo -- visible in the NVSS, at 1.4\,GHz, with a resolution of 45 arcseconds -- can be spotted. This is obviously not the case of LOFAR and GMRT data, where one can study the radio halo of the system in a great detail. An area of fading out emission -- east from the central AGN -- can be easily seen, being the most distinct difference between both of these maps. A detailed analysis of the emission from HCG\,60, including both low-, and high-frequency data from LOFAR, GMRT, and VLA is underway.

HCG\,60 is a good example why additional case studies, and/or a larger statistical one, are still needed. One of the biggest challenges to overcome when identifying radio emitting groups is to differentiate objects where the radio emission originates within the intra-group medium -- thus, is bound to the group as a system -- from those where the extended radio structure is just a smeared out contribution from compact objects, eg. background sources that happen to shine through the group, or a radio galaxy that is inside a group. HCG\,60 is an embedded system, the dense core of the Abell\,1452 cluster. But it is a home to another entity, too: its A member, PGC\,038065, is a wide-angle-tail (WAT) radio galaxy. It has been first described by \citet{jagers87}, who used the WSRT 610\,MHz data to study its morphology. However, for the sake of the groups' magnetism studies, presence of such objects is a serious problem. As already hinted by the map above, the bulk of the radio emission originates within the radio jets and lobes. It is unknown whether the detected structures can be attributed to the intergalactic matter inside the system. This problem might be present in many of the groups that happen to be radio emitters: as said before, they are angularly small, and prone to the beam smearing effect. This further justifies the need to gather data that offer high resolution \emph{and} sensitivity at the same time. 

Last but not least, the coincidence between the neutral gas tails, and non-thermal radio emission is another topic worth exploring. Groups, contrary to clusters, can still host large reservoirs of neutral gas \citep{verdesmontenegro01}, and many examples of impressive, gaseous tails and rings are known (eg. in the Leo Triplet, \citealt{haynes79, stierwalt09}, in the Stephan's Quintet \citealt{williams87,williams02}, or in HCG\,44, \citealt{serra12}). However, there has been little to no success in detecting the continuum counterparts of all these structures. An attempt to find traces of such in the Leo Triplet \citep{bnw13A} at 2.64\,GHz has not been successful; only background sources of radio emission were discovered. A possible trace of an intergalactic structure in HCG\,44 at 4.86\,GHz was reported by \citet{bnw17}, but neither does it coincide with the giant ${\rm H}_{\rm I}$ tail discovered by \citet{serra12}, nor does it reach a comparable extent. 
Probably the best studied example is that of the Stephan's Quintet -- again. The recent study by \citet{bnw20} reveals that the non-thermal, intergalactic structure associated with that group follows the neutral hydrogen tail that protrudes from the southern edge of the system \citep{williams87, williams02}. This structure also encompasses the TDG candidate SQ-B, raising the question whether it is immersed in the main gaseous tail, or in the optical one, the latter being an elongated spiral arm of NGC\,7319. Success of this study confirms that, at least in some cases, a magnetised structure can follow the neutral gas one -- and if only its extent is substantial, then that particular galaxy group is able to participate in the further magnetisation of the intergalactic space. The question is, whether such structures are common (and groups in general can magnetise the Universe), or can be found only in a handful of systems. With the help of the low-frequency instruments, tracing weak radio emission associated with tidal tails is much easier than ever before.

\section{Radio emission from fossil groups}

Fossils are scarcely studied, and not much is known about their magnetic fields. The expectation is that it should be on a downward slope of its evolution, when it comes to its strength: as \citet{drzazga11} argues, the decrease begins after the nuclear coalescence, and fossil is clearly a post-merger system.  However, as already stated, this subclass is likely to contain objects that are and are not relaxed -– what could be reflected in the strength of their magnetic fields, too. So studying the magnetic field, and radio emission could serve as a tool to differentiate object where the "fossilisation" is still ongoing from those where this process has already ended. It should also be easy to assess the importance of magnetic field in the dynamics of a fossil group, by comparing the energy density contained within the magnetic field to that of the thermal plasma -- the latter information extracted from the X-Ray data. This process is more difficult in case of "normal" groups, where in general the X-Ray brightness is lower, and often only an upper constraint for the  gas temperature can be obtained. However, as for now, there is only one galaxy group, which is likely a fossil, and has its radio emission analysed: this is HCG\,15.

HCG 15 is a relatively distant group of six galaxies, four of them being classified as ellipticals and lenticulars. It is not a visually interesting target, with no apparent signs of ongoing, or previous interaction. HCG 15, is however, a very atypical compact group: it is extremely deficient in neutral gas \citep{verdesmontenegro01}, shows large amounts of diffuse, intergalactic light \citep{darocha08}, and has an extended X-Ray halo \citep{mulchaey03}. The presence of the latter was the reason why this group, among 17 others, was chosen for a joint radio-X-Ray study by \citet{giacintucci11} -- aiming at studying the interrelations between the AGN, and the surrounding intra-group medium. Once again, HCG\,15 was found to be an unusual system: not only was it the only system in the study to contain a few bright galaxies, instead of just one, but also the radio emission -- in a form of more than 100-kpc wide halo -- is likely a remnant of a galaxy passing through the group's core. Unfortunately, lack of other radio data prevented further studies.

This system was revisited by \citet{bnw17}, who detected an abrupt fade-out of the radio halo between 1.4 and 4.86 GHz -- something not seen in the other systems analysed by them. This clearly suggested that the radio emission is quite old; also, no obvious signs of diluted radio lobes, or jets of the AGN in HCG\,15A have been found. This altogether suggest that HCG\,15 is indeed an example of a radio emitting fossil. However, it is quite a "lucky" example: although effects of ageing radio emission are easily detectable, the halo remains visible even at 1.4\,GHz. Thus, it is not an extremely old object -- this conclusion is backed by the presence of more than one galaxy  -- and can be still studied using observations from higher frequency instruments. In order to reach the remnants of radio emission in older fossils, a low frequency instrument would be a necessity: radio telescopes like the ILT, GMRT, or SKA are best suited for this kind of studies. 

From the fossils' point of view, the most promising source of the data could be the LOFAR LBA Survey (LoLSS, \citealt{degasperin21}. Designed as a project complementary to the LoTSS, LoLSS features a very similar survey design -- but this time, at 42 -- 66\,MHz, with an expected sensitivity of 1\,mJy per beam (of 15\,arcsec). So far, a preliminary data release has been made, with a lower resolution and sensitivity (47\,arcsecs, and 5\,mJy per beam). Still, these are the first observations with such a high sensitivity and fidelity done below 100\,MHz. In the nearest future, with the subsequent data releases, LoLSS is likely to become an important tool in revealing the oldest radio emission. Thus, an observing window for the oldest fossils will be opened.

\begin{figure*}[htp]
\centering
	\includegraphics[width=0.45\textwidth]{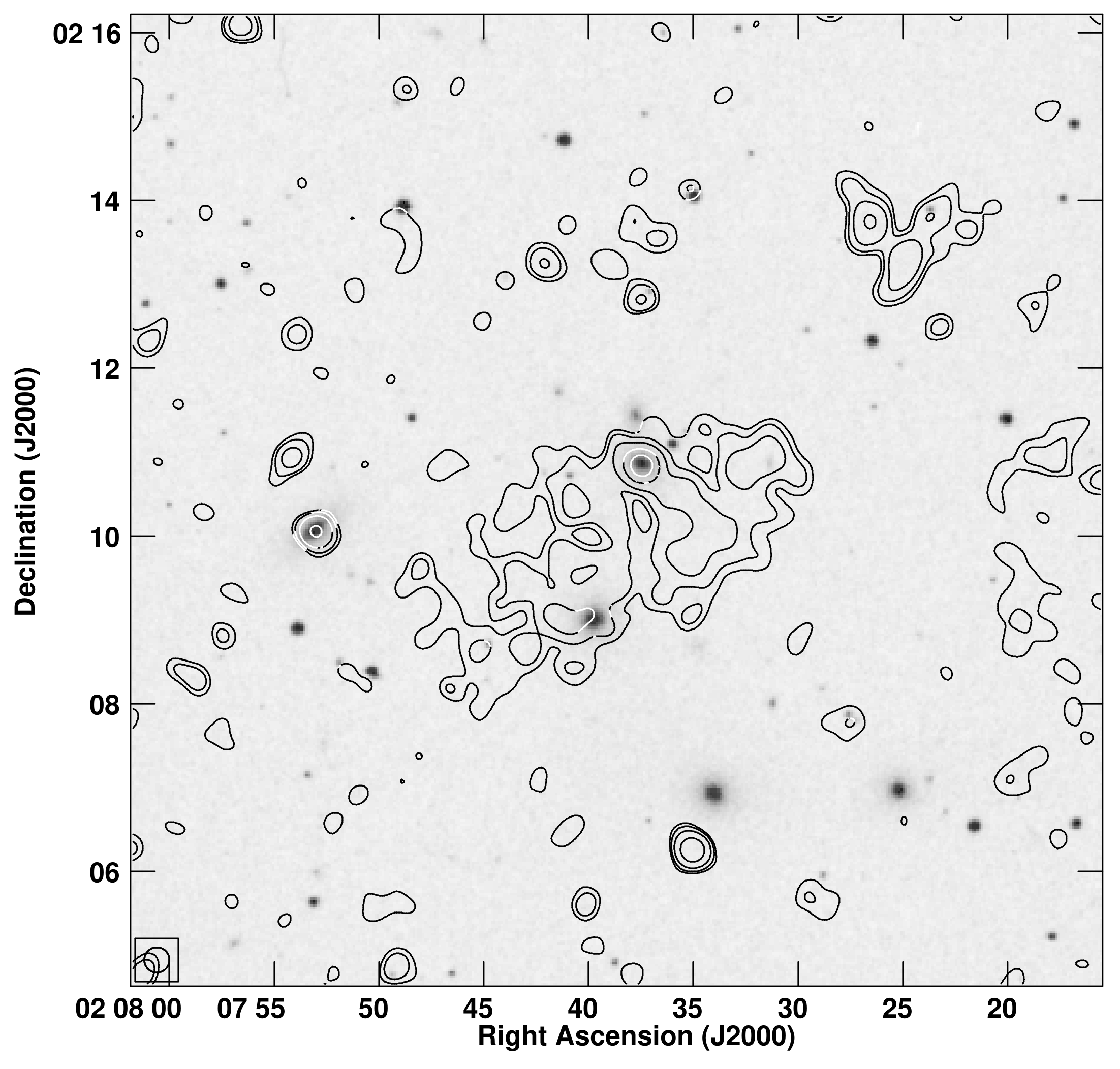}
	\includegraphics[width=0.45\textwidth]{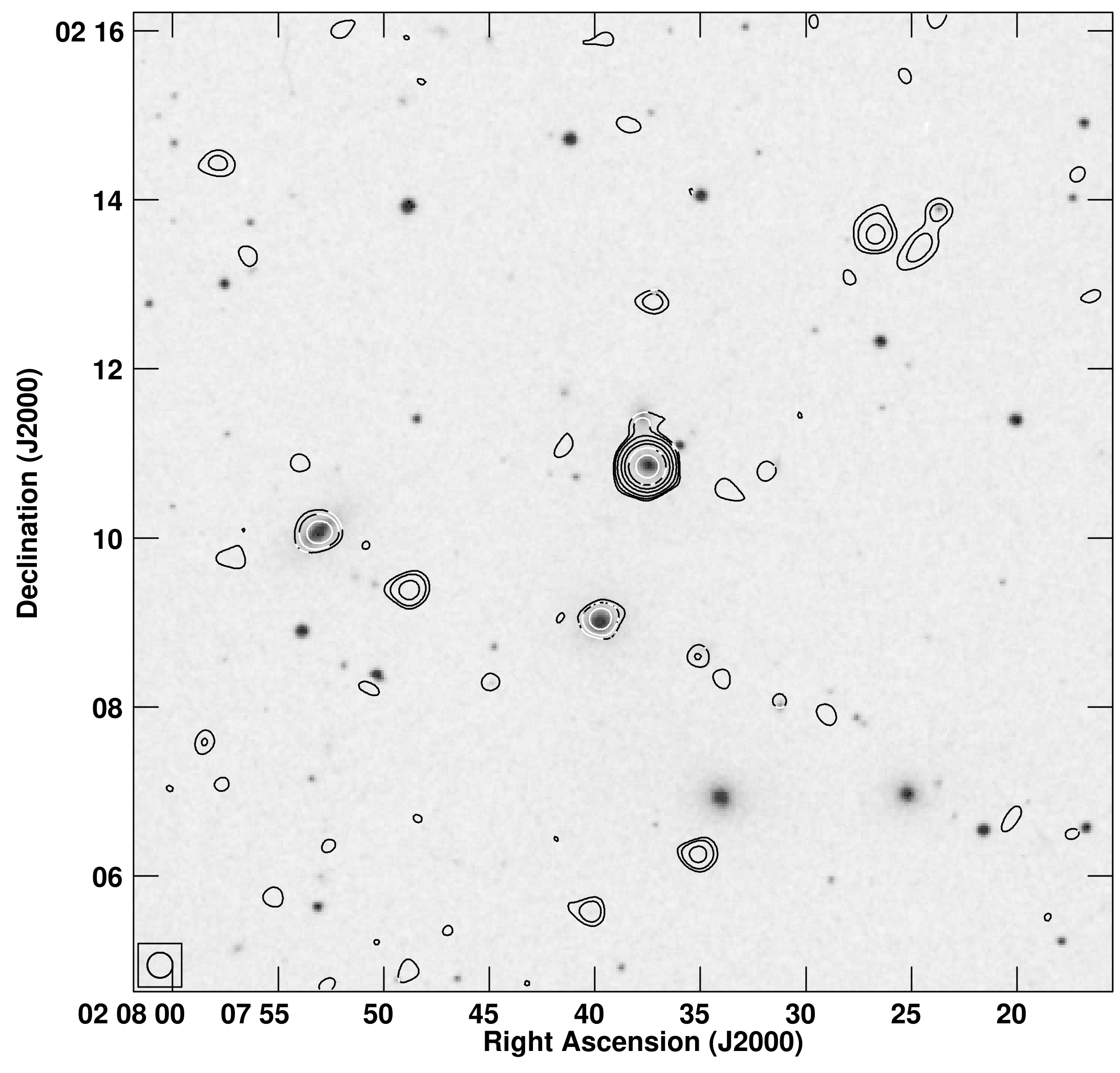}
\caption{\label{fig_h15} Radio emission of HCG\,15 at 618\,MHz (left panel), and at 4860\,MHz (right panel). Radio contours overlaid on a Johnson R filter map from the DSS. The levels are -5, -3, 3, 5, 8, 13, 21, 34, 55, 89, 134, 213, and 345 $\times$ r.m.s. noise level of 45\,$\mu$Jy/beam (618 MHz), and 10\,$\mu$Jy/beam (4860 MHz). The angular resolution of both images is 6 arcseconds.}        
\end{figure*}

Figure ~\ref{fig_h15} shows the aforementioned HCG\,15. Observations at two different frequencies are presented: one at 618\,MHz, one at 4860\,MHz. The difference is striking: while the reprocessed low-frequency data of \citet{giacintucci11} show a large pool of extended radio emission, covering some of the group members, with a global maximum associated with the northernmost galaxy, those gathered at the higher frequency reveal nothing but the point-like radio galaxy. Contrary to the example of HCG\,60 (Fig.~\ref{fig_h60}), neither radio jets, nor lobes are present, suggesting that the system's envelope is indeed a genuine intergalactic structure. Albeit HCG\,15 is a bit atypical for a fossil, as its continuum radio emission can be detected at least as high as 1.4\,GHz \citep{bnw17}, it is only possible to identify and examine the small-scale structures when low-frequency data are used. Being a group of the northern sky, HCG\,15 will finally be included in the LoTSS and LoLSS data. This will allow to perform a detailed study of a fossil's magnetic field for the first time.

What are the scientific objectives that relate to the fossils? In general, as the observational knowledge of fossils in the radio domain is at an even lower level than that of the non-fossil galaxy groups (eg. hardly anything is known about them), the main problem is to characterise the magnetic field of fossils. Do we find radio emission in most of them, or are there any systems with no detection, even at the lowest radio frequencies? How old are the radio haloes in fossils? Metrewave data should make it possible to look at the remnants of processes that ceased more than a hundred Myrs ago. Is it possible to draw a fossil-stage evolutionary sequence using the radio data, and investigate their magnetic fields – so, get a detailed hint into the last dying breath of a galaxy group? Is there any chance to see a fossil being reborn –- a system that is likely to look quite like a uninteresting group, with some weak, but young radio emission at higher frequencies, but has an old halo, seen only at the lowest ones? Are magnetic fields in fossils already weak, as one could expect from the evolutionary sequence of pairs \citep{drzazga11}, or not? Can one differentiate fossils on the basis of the strength of their magnetic field? And what about the galaxies inside: is there any difference between the magnetic fields of field galaxies, and fossil members? All of these questions can only be answered when low frequency, sensitive radio observations are conducted.

While studying fossils, a similar problem to that outlined for "normal" groups is crucial to be addressed: compact sources of radio emission can be smeared out and mimic the intergalactic emission. This could be especially problematic at the lowest radio frequencies, where old, and diffunded lobes of radio galaxies will be prominent. For example, there exists a loose group, immersed in a large, steep-spectrum radio-halo that is interpreted as an old remnant of the AGN activity. At the first glance, NGC 5580/88 \citep{degasperin14} seems to be remarkably similar to HCG 15: albeit it is composed of both early and late type galaxies, it also has a large radio halo, which was once described as a scaled down, cluster-like one. Analysis of the GMRT data reveals the presence of a second, lobe-like structure, suggesting that what is seen in that system is just a remnant of a relatively old AGN activity, not of some merging processes. This further emphasises how beneficial the existing low frequency interferometers -- and the planned ones -- are: only they are able to provide both sensitivity and resolution high enough to allow studying radio emission of fossils. 

\section{In search for the ordered and regular magnetic fields in galaxy groups}
\label{sec_regfield}

Not only turbulent magnetic fields fill the cosmic space. If the magnetic field retains a constant direction over a larger extent, it is classified as an ordered, or a regular one. Such a field manifests through polarised radio emission. Those two genres of the non-turbulent fields differ primarily in the mechanism of their creation. An ordered, but not unidirectional one is created when magnetised gas is either stretched, or compressed. Tidal interactions are therefore likely to produce this genre. For example, in the Stephan's Quintet, existence of an ordered magnetic field has been confirmed by \citep{bnw13B} --  most of the intra-group space is filled with polarised radio emission, including a bridge-like structure between interacting galaxies NGC\,7318A and NGC\,7319. This bridge was not detected beforehand, as it is only visible in the polarised intensity maps.

However, it is not the ordered magnetic field that draws most of the attention -- it is the regular one that does so. The reason is simple: it takes a lot of effort to create such a field. Turbulence, shear, and rotation -- preferably the differential one, although the rigid rotation can also suffice in some cases \citep{siejkowski10, siejkowski14} -- these three (or two) conditions must be fulfilled in order for the MHD dynamo to efficiently amplify and regularise the magnetic field. This process is present in most of the spiral, and many of the irregular galaxies, but not in the elliptical ones \citep{beck13}. Thus, detection of the regular magnetic field signifies an action of the dynamo. However, for many years, it was not easy to differentiate the non-turbulent forms of the magnetic field.

Both regular and ordered magnetic fields manifest in the radio continuum in the same way: by a polarised emission. Vectors misaligned by 180 degrees look the same; any structure detectable in the polarised intensity can be either due to the action of an MHD dynamo (eg. magnetised gas expelled into the intergalactic space, regularised beforehand), or because of some stretching/compressing mechanism (eg. a previously turbulent field, swept by the passage of a shock front). The best way to differentiate these two is to analyse the Faraday effect -- quantified by the Rotation Measure (RM). Those genuinely regular should have a consistent, non-zero RM, those "just" ordered -- the opposite. However, as such a study usually utilises a number of polarised, compact sources that "shine through" the intergalactic regions, it was not an easy job: groups, and pairs are mostly angularly small, and finding enough objects to measure their RM turned out to be almost impossible. Also, to unambiguously determine the RM, observations at at least three different frequencies are needed (as the Faraday effect scales with the wavelength squared). This has all changed thanks to the introduction of the Rotation Measure Synthesis.

The concept of recovering the Faraday information in cases of a non-complete frequency sampling dates back to \citet{burn66}, and was implemented under the name of RM Synthesis by \citet{brentjens05}. The Synthesis requires a dense frequency sampling; the best results are achieved for the wideband data. As a result, this technique works the best for the newest radio telescopes; execution of this procedure on eg. historical VLA datasets is of no use. It is basically a Fourier process, where information in the Stokes Q and U channels (amplitude as a function of frequency) is transformed into the Faraday Space. A sum of contributions from all "slabs" of regular magnetic field encountered on the way from the source to the observer can be disentangled, and detected at different Faraday depths -- and that allows to subtract the amount of rotation introduced by the Milky Way. Since its introduction, this method allowed to study regular magnetic fields from pulsars, through galaxies, to clusters. And, of course, to do the same for pairs and groups.

The first attempt to reveal a regular magnetic field through the RM Synthesis was carried out by \citet{farnes14}, who used the GMRT 610 MHz full polarisation mode to conduct observations of two galaxy groups, namely USGC\,S063, and the Grus Quartet. A very high sensitivity in the Faraday space has been reached; enough to detect emission polarised emission associated with two member galaxies, and three background sources. However, the IGM of both groups turned out to be devoid of regular fields. However, a success was achieved not so long after -- for an intergalactic structure in a pair.

In 2017, successful, spectropolarimetric observations of coherent magnetic fields in the intergalactic spaces of two galaxy pairs have been reported: these were the Antennae system \citep{basu17}, and the Magellanic Clouds \citep{kaczmarek17}. The Antennae are a southern pair of merging spirals best known for their tidally distorted spiral arms, resembling insect's antennae (thus the name). This system is known to host strong magnetic fields, both turbulent, and coherent \citep{chyzy04}. The new study allowed to detect the largest regularly polarised structure at that time: a 20-kpc long section of the tidal tail, possessing a regular magnetic field as strong as 8.5\,$\mu$G. In case of the Magellanic Clouds -- two irregular members of the Local Group -- the detected regular component was weaker, but still significant for the system: the Authors pointed out that the field in question was a probe for a much larger field, not only filling the Magellanic Bridge itself, but rather the whole pair. Following the success of these studies, the Quintet was chosen as a subsequent target.

The Quintet was observed using the WSRT, in two different spectral windows, both within the L-Band \citep{bnw20}. This setup was adopted from a successful spectropolarimetic project on nearby galaxies, the WSRT-SINGS \citep{heald09}, and to ensure the equivalent of angular resolution in the Faraday space -- the Rotation Measure Transfer Function, RMTF -- is as high as possible. The results turned out to be quite unexpected: analysis of the Faraday spectra of sources shining through the Quintet's intragroup region revealed a presence of a large scale "screen" of magnetic field -- most probably the group IGM itself. Analysis of the data could only return the product of its strength and extent along the line of sight; thus, \citet{bnw20} used the previously known characteristics of the system (size, total strength of the magnetic field) to tentatively estimate the strength of the regular component as 2--5\,muG, and its depth as 10--20\,kpc. The most surprising finding was its extent across the sky plane: the regular magnetic field within the group stays coherent within the area of at least 60 on 40 kpc. This is the largest regularly magnetised structure detected on galactic scales. Its origin is puzzling: no mechanism of regularisation is known to work on such a scale, so it must have been created inside one of the group galaxies. 

The fact that spectropolarimetric studies are nearly absent leads to the main objective concerning the presence of regular magnetic fields in galaxy groups. It is necessary to figure out whether this phenomenon is only found in a handful of systems (or possibly, only inside the Quintet?), or -- similarly to the turbulent fields -- might be present in a statistically important fraction of galaxy groups. Studies on galaxy pairs suggest that a regular magnetic field can dominate over the turbulent one -- at least locally. This is another question worth answering in case of groups -- can the situation be the same here, or rather the regular one is but a small add-on to the turbulent fields? Instruments like the JVLA, or uGMRT -- upgraded with wide-band correlators -- already offer superb capabilities for spectropolarimetric studies. The expected SNR achieved by those radio telescopes should be more than enough to provide data that can be subjected to the RM Synthesis. LOFAR is also capable of this kind of studies. Attempts to find the regular magnetic field signatures have already been carried out for HCG\,60, during the MKSP Busy Weeks in 2020. However, the results are still ambiguous, as the detected polarised signal is weak, and no magnetic field coherent over larger scales was revealed. 

The forthcoming years will also add another instrument to this puzzle. The SKA is expected to start operations in this decade. One of the goals of this instrument, pursued by its Cosmic Magnetism Science Working Group, is to study the cosmic magnetic field. Within this framework, projects featuring a deep analysis of the magnetic field inside nearby galaxies have been proposed: one that utilises the SKA Band 4 (2.8 -- 5.2 \,GHz, \citealt{beck20}, and one that uses the SKA Band 2 (1.0 -- 1.8\,GHz, \citealt{heald20B}). The expected features of these studies -- accessible Faraday depth, resolution in the Faraday space, and angular resolution of a few arcseconds should be more than enough to conduct studies on regular magnetic fields in galaxy groups.

There is probably not much hope for detecting this feature in fossils -- or in, general, in groups at later evolutionary stages, as the fragile structure of the intergalactic, regular magnetic field can easily be tangled or ripped during the subsequent interactions. Less dense, "alive" environments, on the other hand, are probably more prone (maybe even more than the compact ones?) for the magnetic fields to pertain relatively unscathed in their IGM. 
Large gaseous outflows, like the tail of the Leo's Triplet \citep{haynes79,stierwalt09}, or the neutral gas ring circumventing HCG\,44 \citep{serra12}, mentioned already in Sect.~\ref{sec_general}, are another promising target for the spectropolarimetric studies. When observed with both high resolution and sensitivity, they should reveal enough polarised background sources "shining through" the tidal features to search for the signatures of a large-scale, regular magnetic field that introduces additional Faraday Rotation to the original signal.

\section{Concluding remarks}

Groups are the most widespread galactic environment in the Universe -- and possibly the most special one. They are much more complex systems than pairs, while simultaneously they clearly do not follow the self-scaling laws defined for clusters. Dense environment, and the frequency of interactions result in an accelerated evolution of member galaxies -- which is reflected in their star-forming properties. While the interest in studying groups has largely grown since the early works of \citet{shakhbazyan73} and \citet{hickson82}, not much has been said about their magnetic fields, and the radio emission through which this phenomenon manifests -- despite its importance for the galaxies themselves. The following points are recapitulating the current state of the knowledge, and list out the main goals for future studies -- studies that have been made available thanks to the introduction of the new generation of radio interferometers.

\begin{itemize}
    \item More than 15\% of galaxy groups are likely to host detectable radio emission, and their IGM is expected to host magnetic fields of comparable strength to those found inside spiral galaxies. The majority of groups contain at least one -- likely more -- radio-emitting galaxy;
    \item Reliable detection of a radio-emitting fossil, HCG\,15, has also been made. This system hosts a detectable magnetic field, filling a 100-kpc wide intergalactic halo, in which the group is immersed. No traces of radio jets or lobes have been found in the system, confirming the non-AGN character of the emission;
    \item Spectropolarimetric capabilities of the wide-band, multi-channels correlators allowed to detect the presence of a genuinely regular magnetic field inside the Stephan's Quintet. This the largest known, regularly magnetised structure detected on a galactic scale, surpassing any of the previously known objects of this type -- and is likely even larger, as only lower constraints for its size could be determined;
    \item It is necessary now to expand the studies on groups' magnetism into an even larger sample that will allow to execute statistical operations on these objects, to compare galaxies within and outside groups, and map the magnetic field's strength inside them;
    \item New observations should also make it possible to reliably sort out objects where most of the radio emission is associated with AGN, not with the system's halo -- like in case of HCG\,60, where high fidelity data suggests that old, diluted radio lobes comprise most of the detected structure;
    \item With the help of the low frequency radio observations, detection of emission from the oldest fossils should be possible -- especially, as data of superb quality will soon be available at frequencies well below 100\,MHz. These data will be crucial to assess the question whether the magnetic in groups follow similar evolutionary track as those found in pairs, or not. 
    \item Should there exist a well-defined track of the magnetic fields' evolution, the phenomenon in question will then become a useful tool in finding out the evolutionary stage of given objects;
    \item It is also possible to examine whether the regular magnetic field found inside the Stephan's Quintet is a mere exception, or rather a typical phenomenon. In particular, the spectropolarimetric observations can be a fruitful way of searching for regular magnetic field in large tidal structures.
    
\end{itemize}
\vspace{6pt} 

\funding{This work has received support from the Polish National Science Centre (NCN), grant no. UMO-2016/23/D/ST9/00386.}

\institutionalreview{Not applicable}

\informedconsent{Not applicable}

\dataavailability{Data used to describe the exemplary galaxy groups was taken from the GMRT Online Archive, and from the NRAO Science Data Archive} 

\conflictsofinterest{The author declares no conflict of interest. The funders had no role in the design of the study; in the collection, analyses, or interpretation of data; in the writing of the manuscript, or in the decision to publish the results.} 

\acknowledgments{

The data used in this work was in part processed on the Dutch national e-infrastructure with the support of SURF Cooperative through grant e-infra 160022 \& 160152. This paper is based (in part) on data obtained with the International LOFAR Telescope (ILT). LOFAR (van Haarlem et al. 2013) is the Low Frequency Array designed and constructed by ASTRON. It has observing, data processing, and data storage facilities in several countries, that are owned by various parties (each with their own funding sources), and that are collectively operated by the ILT foundation under a joint scientific policy. The ILT resources have benefited from the following recent major funding sources: CNRS-INSU, Observatoire de Paris and Université d'Orléans, France; BMBF, MIWF-NRW, MPG, Germany; Science Foundation Ireland (SFI), Department of Business, Enterprise and Innovation (DBEI), Ireland; NWO, The Netherlands; The Science and Technology Facilities Council, UK; Ministry of Science and Higher Education, Poland. The author thanks the Ministry of Science and Higher Education (MSHE), Poland for granting funds for the Polish contribution to the ILT (MSHE decision no. DIR/WK/2016/2017/05-1)" and for maintenance of the LOFAR LOFAR PL-611 Lazy, LOFAR PL-612 Baldy stations (MSHE decisions: no. 46/E-383/SPUB/SP/2019 and no. 59/E-383/SPUB/SP/2019.1, respectively).
I thank the staff of the GMRT that made these observations possible. GMRT is run by the National Centre for Radio Astrophysics of the Tata Institute of Fundamental Research.
The National Radio Astronomy Observatory is a facility of the National Science Foundation operated under cooperative agreement by Associated Universities, Inc.
}

\abbreviations{The following abbreviations are used in this manuscript:\\

\noindent 
\begin{tabular}{@{}ll}
AGN   & Active Galactic Nuclei\\
AIPS  & Astronomical Image Processing System \\
FIR   & Far InfraRed\\
FIRST & Faint Images of the Radio Sky at Twenty centimeters\\
GMRT  & Giant Metrewave Radio Telescope \\
HCG   & Hickson Compact Group\\
HETDEX& Hobby-Eberly Telescope Dark Energy eXperiment\\
IGM   & InterGalactic Medium \\
ILT   & International LOFAR (LOw Frequency Array) Telescope \\
LBA   & Low Band Antenna\\
LOFAR & LOw Frequency ARray\\
LoLSS & Lofar Lba Survey \\
LoTSS & Lofar Two-metre Sky Survey\\
MDPI  & Multidisciplinary Digital Publishing Institute\\
MHD   & MagnetoHydroDynamics\\
MKSP  & Magnetism Key Science Project\\
MWA   & Murchinson Widefield Array\\
NGC   & New General Catalogue\\
NRAO  & National Radio Astronomical Observatory\\
NVSS  & NRAO VLA Sky Survey\\
PGC   & Principal Galaxies Catalogue\\
POSS  & Palomar Optical Sky Survey \\
RM    & Rotation Measure\\
RMTF  & Rotation Measure Transfer Function\\
SINGS & Spitzer Infrared Nearby Galaxies Survey\\
SKA   & Square Kilometre Array\\
SNR   & Signal to Noise Ratio\\
SPAM  & Source Processing and Atmospherical Modelling\\
TDG   & Tidal Dwarf Galaxy\\
TIFR  & Tata Institute of Fundamental Research\\
uGMRT & upgraded GMRT\\
USGC  & Updated Zwicky--Southern Sky Redshift Survey Group Catalog\\
VLA   & Very Large Array\\
WAT   & Wide-Angle Tail\\
WSRT  & Westerbork Synthesis Radio Telescope
\end{tabular}}

\end{paracol}

\reftitle{References}

\end{document}